\documentclass[twocolumn,prl,showpacs]{revtex4}  

\usepackage{graphicx}
\usepackage{amssymb}

\begin{document}

\title{Universality of liquid dynamics}

\author{T. Oppelstrup $^{1,2}$, B. Sadigh $^{2}$, S. Sastry
$^{3}$, and M. Dzugutov $^{4}$}

\affiliation{$^1$ Dept. for Numerical Analysis and Computer Science,
Royal Institute of Technology, S--100 44 Stockholm, Sweden\\ $^2$
Lawrence Livermore National Laboratory, Livermore, CA 94550, USA\\
$^3$ Javaharlal Nehru Centre for Advanced Scientific Research, Jakkur
Campus, Bangalore 560064, India \\ $^4$ Dept. for Materials Science
and Engineering, Royal Institute of Technology, S--100 44 Stockholm,
Sweden}

\begin{abstract}

We investigate the origin of the Stokes-Einstein relation in
liquids. The hard-sphere dynamics is analyzed using a new measure of
structural relaxation - the minimum Euclidean distance between
configurations of particles. It is shown that the universal relation
between the structural relaxation and diffusion in liquids is caused
by the existence of one dominating length scale imposed by the
structural correlations and associated with de Gennes narrowing. We
demonstrate that this relation can be described by a model of
independent random walkers under the single-occupancy constraint.
\end{abstract}
\date{September 12, 2005}

\maketitle

Much effort in the studies of liquids has been focused on
understanding the relationship between the ergodicity-restoring
structural relaxation and diffusive motions of individual particles
\cite{hansen, yip}. It has been recognized that, at least in simple
liquids, the self-diffusion coefficient $D$ can be derived from static
correlation functions without any knowledge of the interparticle
forces \cite{MD, babak}. On the other hand, kinetic theories
\cite{Gotze92, young} have been successful in expressing
time-dependent correlation functions describing the relaxation
dynamics in liquids in terms of structural correlations. These results
suggest a purely structural mechanism linking the two aspects of
liquid dynamics. The existence of such link can also be inferred from
the Stokes-Einstein \cite{einstein} and Stokes-Einstein-Debye
\cite{debye} relations which can be combined in the following relation
\cite{landau, tarjus}:
\begin{equation} 
 D \tau \sim a^2
\end{equation} 
where $\tau$ is the characteristic time of structural relaxation, and
$a$ is the size of the diffusing particle. We note that in dense
fluids, $a$ can be interpreted as the close-neighbour distance which
represents the dominating length scale of the structural correlations.

The current interest in the mechanism of the Stokes-Einstein relation
is motivated by the recent finding that this relation breaks in
strongly supercooled fragile glass-forming liquids \cite{cicerone,
Ediger96,Chang}. A number of models have been proposed to interpret
this observation \cite{tarjus,stillinger}. However, we must first
understand why these relations, derived from purely hydrodynamic
postulates, demonstrate such remarkably universal success in
describing the atomic-scale dynamics in the normal liquid state.

In this Letter, we analyze the origin of the Stokes-Einstein dynamics
in liquids. The dynamics of hard spheres (HS) is examined using a new
measure of structural relaxation based on the assessment of minimum
distance between the system's configurations. We present evidence that
the invariance of the relaxation rate relative to the diffusion rate
in the liquid state is linked to the existence of a single
structure-induced length scale dominating the liquid dynamics which is
manifested by de Gennes narrowing \cite{degennes}, and it also implies
Gaussianity of the diffusion process. We show that this dynamics can
be described by a simple model of independent random walkers confined
to single-occupancy cells \cite{hoover, kirkwood}.

The dynamics of a system of identical particles is commonly described
in terms of the van Hove density correlation function $G(r,t)$
\cite{hansen}. Its self-part, $G_s(r,t)$ describes the evolution of
the density distribution for a tagged particle. However, because the
particles are indistinguishable, the evolution of a tagged particle
becomes irrelevant for the structural relaxation as soon as the
particle traverses the nearest neighbour distance.

For a system of $N$ identical particles, there exists $N!$ physically
indistinguishable permutations of each configuration. Let $ {\bf r}^N
(t) = \{ {\bf r}_i(t) \} $ be the system's configuration at a time
moment $t$. We define a correlation function:
\begin{equation} 
 G_n(r,t)= \frac{1}{N}\left \langle \sum_{i=1}^N \delta \left [ r -
 \left |{\bf r}_i(0) - {\bf r}_{p(i)}(t) \right | \right ] \right
 \rangle
\end{equation} 
$ \{p(i) \} = \hat{P} \{ i \}$, where $\hat{P}$ is the permutation
 procedure \cite{permutation} minimizing the second moment:
\begin{equation} 
 d(t) = \int_0^{\infty} r^2 G_n(r,t) 4 \pi r^2 d r = \frac{1}{N}
\sum_{i=1}^N \left [ {\bf r}_i (0) - {\bf r}_{p(i)} (t)\right ]^2
\end{equation} 
This quantity, to be referred to as {\it dissimilarity}, represents
the minimum square Euclidean distance between the two
configurations. In general, this minimizing relabeling procedure
($d$-mapping) can be used to evaluate the dissimilarity between any
two configurations:
\begin{equation} 
 d( {\bf a}^N, {\bf b}^N) = \frac{1}{N} \sum_{i=1}^N \left [ {\bf a}_i
 - {\bf b}_{p(i)} \right ]^2
\end{equation} 

We use this approach to analyze a molecular dynamics model of 27000
identical HS particles. Throughout this study, the length is measured
in terms of $\rho^{-1/3}$, $\rho$ being the number density.

\begin{figure} 
\includegraphics[width=6.2cm]{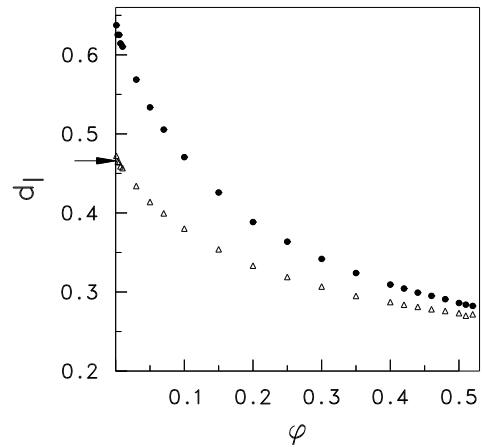}
\caption{ Dots, $d_l$ for two uncorrelated HS fluid configurations;
triangles, $d_l$ for a HS fluid configuration relative to the fcc
reference configuration. The arrow indicates $d_l$ for perfect gas
constrained to the single-occupancy fcc Wigner-Seitz cells
\protect\cite{hoover, kirkwood} multiplied by $e^{2/3}$. }
\label{fig:1}
\end{figure}

For high packing fractions and short time intervals where the number
of relabellings is negligible, $G_n(r,t)$ would behave as $G_s(r,t)$.
Therefore, the former, like the latter, is expected to have a Gaussian
form. However, unlike $G_s(r,t)$ the second moment of which - the mean
square displacement (MSD) - grows linearly with time, $G_n(r,t)$
converges to a limit distribution with a finite second moment
$d(t)\rightarrow d_l$ as the two configurations become uncorrelated
for $ t\rightarrow \infty$ . The variation of $d_l$ in the HS system
is plotted in Fig.~\ref{fig:1} as a function of the packing fraction $\varphi$.

For a selected reference configuration, the described permutation
$\hat{P} $ uniquely maps the system's configuration space onto a
compact region possessing the following properties:

\noindent
(i) For $N\rightarrow \infty$ almost all its points are infinitely
close to the surface of a sphere of square radius $Nd_l$ \cite{Ma}
centered at the reference configuration ($d$-sphere). Therefore, any
configuration uncorrelated with the reference configuration is
expected to be separated from the latter by $d=d_l$.

\noindent
(ii) $d_l$ separating two uncorrelated configurations is a {\it static
} property entirely determined by the density fluctuations in these
configurations. In the HS system, it increases monotonously with
decreasing packing fraction $\varphi$, see Fig.~\ref{fig:1}. The impact of the
structural fluctuations on $d_l$ can be assessed by using a regular
lattice as a reference configuration. It is shown in Fig.~\ref{fig:1} that for
the entire range of $\varphi$, $d_l$ for the HS system calculated with
respect to the fcc lattice is reduced compared with $d_l$ calculated
for two uncorrelated HS fluid configurations.

\noindent
(iii) In the case of the perfect gas (PG), the volume of the $d$-sphere
scales as the total number of distinguishable configurations. To test
this, we consider a PG model where particles are confined to
single-occupancy Wigner-Seitz cells of the fcc lattice \cite{hoover,
kirkwood}. Its $d_l$ relative to the fcc configuration is equal to the
inertia moment of the cell. The single-occupancy constraint reduces
the total number of PG configurations by the factor $e^N$
\cite{kirkwood}; respectively, $d_l$ is expected to be reduced by the
factor $e^{2/3}$. As shown in Fig.~\ref{fig:1}, this is in good agreement with
the simulation results.

\begin{figure} 
\includegraphics[width=6.cm]{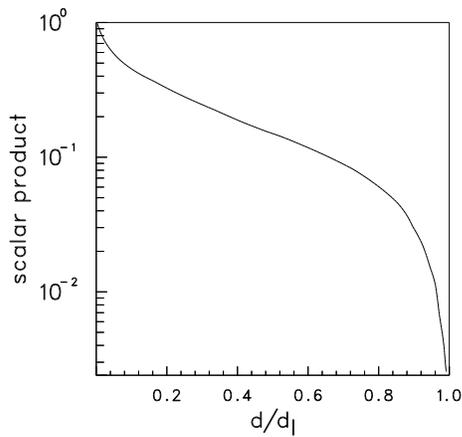}
\caption{ The evolution of the scalar product in the l.h.s. of
Eq.~(5), as a function of $d(t)/d_l$, calculated for the HS liquid for
$\varphi=0.48$. The reference configuration ${\bf c}^N$ is the fcc
lattice.}
\label{fig:2}
\end{figure}

\noindent
We now prove that configurations ${\bf r}^N(0)$ and ${\bf r}^N(t)$ are
uncorrelated if $d(t)=d_l$. Consider $d$-mapping of ${\bf r}^N(t)$
onto the $d$-sphere of a reference configuration ${\bf c}^N$ such that
all its particles have identical environments. Relative to a pole of a
multidimensional sphere, almost all the points of the sphere are
infinitely close to the respective equator. Therefore, if $d(t)=d_l$,
${\bf r}^N(t)$ would be mapped on the equator assuming that ${\bf
r}^N(0)$ is mapped as a pole. The orthogonality condition follows:
\begin{equation} 
 \sum^N_{i=1} [{\bf r}_{p(i)}(t)-{\bf c}_i][{\bf r}_{q(i)}(0)-{\bf c}_i] = 0
\end{equation} 
where $ p(i)$ and $ q(i)$ are permutations $d$-mapping ${\bf r}^N(0)$
and ${\bf r}^N(t)$ onto ${\bf c}^N$. Covariance of two stochastic
variables is a measure of their linear correlation; therefore, (5)
implies that ${\bf r}^N(0)$ and ${\bf r}^N(t)$ are uncorrelated
(notice that $d$-mapping can induce correlation between these
configurations if ${\bf c}^N$ includes any density
fluctuations). Fig.~\ref{fig:2} shows the evolution of the l.h.s. of Eq.~(5) in
the HS liquid for $\varphi=0.48$ as a function of $d(t)/d_l$, the
reference configuration ${\bf c}^N$ being the fcc lattice.

Thus, $d(t)=d_l$ is a necessary and sufficient condition for the two
configurations separated by $d(t)$ to be uncorrelated. $d(t)$ can
therefore be used as a convenient indicator of the system's approach
to its ergodic equilibrium. We propose to measure the
structural relaxation in a system of particles in terms of the
following dimensionless quantity:
\begin{equation} 
 \Phi_d(t) = [d_l-d(t)]/d_l 
\end{equation} 

Structural relaxation in liquids is commonly assessed in terms of the
intermediate scattering function $F(Q_m,t)$, $Q_m$ being the position
of the main maximum of the structure factor $S(Q_m) \equiv F(Q_m,0)$
\cite{hansen, yip, degennes}. For liquid densities, $Q_m$ represents
the length scale of the slowest relaxing structural correlation (De
Gennes narrowing \cite{degennes}). This effect, however, disappears at
low densities \cite{hansen, yip, degennes}. An important advantage of
$\Phi_d(t)$ as a measure of relaxation is that it is not associated
with a particular length scale. Also, unlike the energy-based measure
of ergodic convergence \cite{Mountain89}, its calculation doesn't
require knowledge of the interparticle interaction. Another advantage
of $\Phi_d(t)$ is that, being based on a single-particle correlation
function, it can provide insight into the link between the
individual particle dynamics and the structural relaxation.

\begin{figure} 
\includegraphics[width=8.4cm]{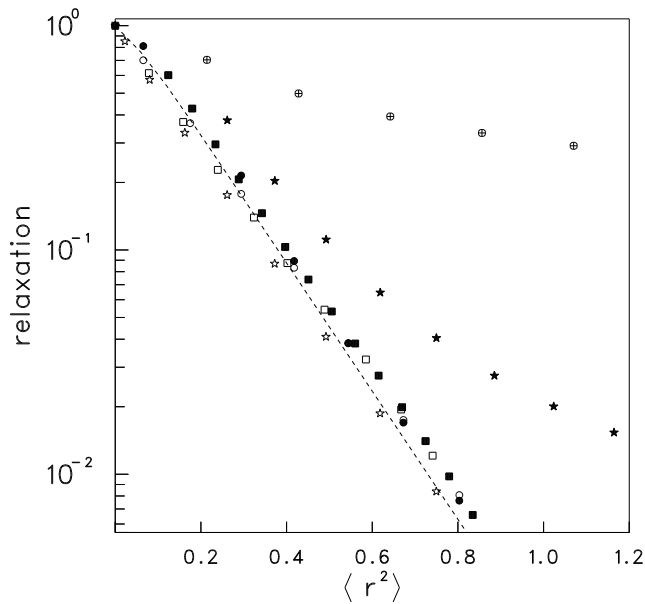}
\caption{Relaxation in the HS system as a function of MSD. Dots,
filled boxes and filled stars: $\Phi_d(t)$ for $\varphi=0.48$,
$\varphi=0.35$ and $\varphi=0.2$, respectively. Open circles, open
boxes and open stars: $F(Q_m,t)/S(Q_m)$ for $\varphi=0.48$,
$\varphi=0.35$ and $\varphi=0.2$, respectively. Crossed circles,
$\Phi_d(t)$ for perfect gas. Dashed line, $\Phi_d(t)$ calculated
according to Eq.~(7) }
\label{fig:3}
\end{figure}

\begin{figure} 
\includegraphics[width=6.cm]{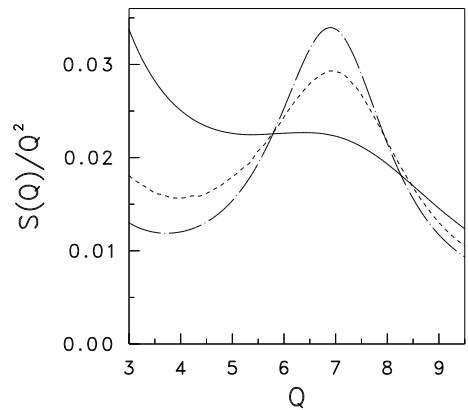}
\caption{ $S(Q)/Q^2$ for the HS fluid representing the $Q$-dependence
of the relaxation time of the intermediate scattering function
$F(Q,t)$ \cite{balucani}. Solid line, $\varphi=0.2$; dashed line,
$\varphi=0.3$; chain-dashed line, $\varphi=0.35$.}
\label{fig:4}
\end{figure}

In Fig.~\ref{fig:3} we analyze the relation between the structural relaxation
and diffusion in the HS system. The structural relaxation is measured in
terms of $\Phi_d(t)$ and presented as a function of MSD. For two
representative values of the packing fraction, $\varphi=0.35$ and
$\varphi=0.48$ which span the liquid domain, the evolution of
$\Phi_d(t)$ is compared with the respective evolution of
$F(Q_m,t)/S(Q_m)$. It is evident that for these two values of
$\varphi$, the decay of $\Phi_d(t)$ is exponential, and it agrees quite
well with the respective behaviour of $F(Q_m,t)/S(Q_m)$. It is also
evident that, within the liquid domain, the rate of structural
relaxation relative to the diffusion rate is invariant with respect to
the variation of $\varphi$.

Fig.~\ref{fig:3} shows that the universal relation between the structural
relaxation and diffusion observed within the liquid domain breaks for
lower densities. At $\varphi=0.2$, we find a distinctly different
regime. The structural relaxation relative to the diffusion as
measured by $\Phi_d(t)$ becomes retarded as compared with the
universal pattern observed for the liquid. 
Notice that the decay of $F(Q_m,t)/S(Q_m)$ for $\varphi=0.2$ agrees
quite well with the exponential liquid pattern. One
can also see that $\Phi_d(t)$ exhibits a strong stretching as compared
with the exponential decay. Both the retardation and the degree of
stretching progressively increase with decreasing $\varphi$.
These results can be compared with an earlier simulation
study of the HS viscosity \cite{alder} where it was found that the
Stokes-Einstein relation breaks around $\varphi= 0.15$.  We note that
the Enskog theory \cite{chapman} predicts violation of the SE relation
in the low-density gas limit, with $D\eta$ diverging as
$\varphi^{-1}$.

These results thus pose us with the following questions: what is the
origin of the universality of the liquid relaxation dynamics and why
does it break in the low-density domain?

The observation that, within the liquid domain of $\varphi$, $\Phi_d
(t)$ decays exponentially relative to MSD implies that the liquid
relaxation is controlled by a single length scale, and the close
agreement between the decay of $\Phi_d(t)$ and the decay of the
respective $F(Q_m,t)/S(Q_m)$ clearly indicates that this length scale
corresponds to the density fluctuation that gives rise to the main
peak of $S(Q)$. This length scale, $2 \pi /Q_m$, is the size of the
first shell of neighbours, the dominant element of the liquid
structure. The maximum of the relaxation time of $F(Q,t)$ at $Q_m$ (De
Gennes narrowing \cite{degennes}) is caused by the so-called ``cage
effect'' \cite{MD, cohen} whereby the structural relaxation is
facilitated by the diffusive motions of the particles within the
confining cages of the nearest neighbours. The cage size thus controls
the relation between the diffusion and the structural relaxation, and
its $\varphi$-invariance explains the latter's universality.

The relaxation time for the $Q$-component of $F(Q,t)$ scales as
$\propto S(Q)/Q^2$ \cite{degennes, balucani, pusey}. Its variation
with respect to $Q$ and $\varphi$ is shown in Fig.~\ref{fig:4}. It features a
pronounced maximum at $Q_m$ for $\varphi=0.35$ where the relaxation
and diffusion follow the universal liquid relation, see Fig.~\ref{fig:3}. For
$\varphi=0.2$, that maximum disappears, and the decay of $F(Q,t)$ is
dominated by the density fluctuations with length scales exceeding
$2\pi/Q_m$. The respective increase in the range of the diffusive
motions dissipating these fluctuations results in the retardation of
the relaxation process relative to diffusion as compared with the
universal liquid behaviour, see Fig.~\ref{fig:3}. Moreover, the stretching of
$\Phi_d(t)$ beyond a single-exponential behaviour can be interpreted
as a superposition of multiple exponential relaxation processes with a
wide range of relaxation rates.

We thus conclude that the observed universality of the liquid
relaxation behaviour relative to diffusion is explained by the
confinement of the relevant density fluctuations to the first
coordination shell which manifests itself as the maximum of the
$Q$-dependent relaxation time at $Q_m$ (de Gennes narowing).

We support this picture of liquid dynamics by the following model.
Consider a tessellation of space into unit-volume cubic cells.
Each cell contains one point particle performing Fickian diffusion
with the self-diffusion coefficient $D$. The diffusion starts from the
centre of the respective cube, and it is confined to the latter by
reflective boundaries. This dynamics relaxes the initial lattice
configuration to the equilibrium perfect gas where the density
fluctuations are constrained by the single-occupancy confinement. This
is supposed to model the confinement of the density fluctuations to
the first neighbour shell that was concluded for the liquid
dynamics. The relaxation of the initial structure is assessed in terms
of $d(t)$. We note that, in this model, reflecting boxes' boundaries
are equivalent to applying the $d$-minimizing permutation, and, for
the uniform equilibrium distribution of particles' density within the
cube, $d_l=1/4$. The solution for $d(t)$ can be obtained as the
following serial expansion:
\begin{equation} 
d(t) = \frac{1}{4} + 6\sum_{k=1}^\infty
\frac{1}{k^2\pi^2}\cos\left(\frac{k\pi}{2}\right)
\left(1+\cos k\pi\right)e^{-k^2\pi^2Dt}
\end{equation} 
The decay of $\Phi_d (t)$ with respect to $\langle r^2 (t) \rangle =
6Dt$ shown in Fig.~\ref{fig:3} demonstrates good agreement with the liquid
relaxation data.

These results highlight two principle aspects of the liquid dynamics
representing essential conditions for the universal relation between
the relaxation and diffusion, and the Stokes-Einstein relation. First,
the relaxation process is dominated by the dissipation of the density
fluctuations confined to the first coordination shell. This behaviour,
also observed as de Gennes narrowing, can be diagnosed from the
structure data as a maximum at $Q_m$ in $S(Q)/Q^2$. Second, the
trajectory of a diffusing particle facilitating the relaxation process
is expected to be fully randomized within the first-neighbour
distance, which implies Gaussianity of the diffusion process.  The
deviation from the universal pattern of liquid dynamics observed in
the low-density domain is evidently caused by the presence of
long-range density fluctuations violating the first of the above
conditions.

S.S. and M.D. gratefully acknowledge support by the Swedish Science
Research Council (VR).

\end{document}